\RequirePackage[utf8]{inputenc}
\RequirePackage[T1]{fontenc}
\documentclass[reprint,pra,superscriptaddress,floatfix]{revtex4-1}
\pdfoutput=1

\usepackage{mathtools}
\usepackage{braket}
\usepackage{bbold}

\usepackage[normalem]{ulem}

\usepackage[single]{acro}
\DeclareMathOperator{\tr}{tr}

\ExplSyntaxOn
\NewDocumentCommand \IncludeGraphicsOrInputTikZ { O{} m }{
	\includegraphics[ #1 ]{ #2 }
}
\ExplSyntaxOff

\usepackage[colorlinks=true,citecolor=blue,linkcolor=red,urlcolor=blue]{hyperref}

\begin{document}
\title{Towards a Spectroscopic Protocol for Unambiguous Detection of Quantum Coherence in Excitonic Energy Transport}

\author{Max Marcus}
\email{max.marcus@chem.ox.ac.uk}
\affiliation{Department of Physics, University of Warwick, Coventry, CV4 7AL, United Kingdom}
\affiliation{Department of Chemistry, Physical \& Theoretical Chemistry Laboratory, University of Oxford, South Parks Road, Oxford, United Kingdom}

\author{George Knee}
\affiliation{Department of Physics, University of Warwick, Coventry, CV4 7AL, United Kingdom}

\author{Animesh Datta}
\email{animesh.datta@warwick.ac.uk}
\affiliation{Department of Physics, University of Warwick, Coventry, CV4 7AL, United Kingdom}

\date{\today}

\begin{abstract}
	The role of quantum effects in excitonic energy transport (EET) has been scrutinised intensely and with increasingly sophisticated experimental techniques. This increased complexity requires invoking correspondingly elaborate models to fit spectroscopic data before molecular parameters can be extracted. Possible quantum effects in EET can then be studied, but the conclusions are strongly contingent on the efficacy of the fitting and the accuracy of the model. To circumvent this challenge, we propose a witness for quantum coherence in EET that can be extracted directly from two-pulse pump-probe spectroscopy experimental data. We provide simulations to judge the feasibility of our approach. Somewhat counterintuitively, our protocol does not probe quantum coherence directly, but only indirectly through its implicit deletion. It allows for classical models with no quantum coherence to be decisively ruled out.
\end{abstract}

\maketitle

\section{Introduction}
The nature of energy transport in light-harvesting complexes, especially the role of quantum effects, is a topic of intense debate~\cite{Engel:2007aa,Plenio:2008aa,Mohseni:2008aa,Caruso:2009aa,Wilde:2010aa,OReilly:2014aa}. Central to this is the excited state dynamics of multichromophoric systems. These dynamics are usually studied with non-linear ultrafast optical spectroscopy, allowing for the investigation of the underlying processes in exciton energy transport (EET). Two-dimensional electronic spectroscopy (2DES) is sufficient to perform full tomography of the quantum process of excitonic dynamics within the singly-excited manifold of light-harvesting complexes~\cite{Yuen-Zhou:2011aa} and determine the precise values of the populations - as well as any quantum coherences between energy eigenstates~\cite{Yuen-Zhou:2011ab,Yuen-Zhou2014,Chenu:2015aa,Pachon2015,Gururangan2019} at the end of a dynamical process transforming any initial excited state. Unfortunately, it is known that this reconstruction can be non-unique~\cite{Ruckebusch2012,Gururangan2019}. The question of the existence of quantum coherences, generated by a process acting on a particular input state is simpler and therefore potentially easier to answer. 

Our objective is to explore the possibility of a simpler spectroscopic scheme that is strictly easier than full process tomography and can still provide unambiguous evidence of quantum coherence. This work thus represents a step into the area of identifying quantum coherences in light-harvesting complexes, which is of fundamental interest but fraught with great difficulties both conceptual and practical. We hope to clarify some of these here.

While non-linear ultrafast spectroscopy experiments routinely give qualitative insights into the complicated nature of the relevant processes~\footnote{Sometimes called `blobology'~\cite{Richter:2018aa}}, quantitative results are more difficult to obtain. In fact, the latter typically require theoretical modelling of the underlying physics~~\cite{Ishizaki2011,Kreisbeck2011} and fitting such simulations to experiment then allows for the estimation of molecular parameters~\cite{Butkus2017}. Although being a workable approach, the complicated nature of ultrafast spectroscopic experiments and the multitude of model parameters may allow for several possible interpretations. In general, this problem becomes more severe as the  experimental set-up of the spectroscopic experiment becomes more complicated.

Identifying quantum coherences in EET is difficult because of the major challenge of constructing tractable theoretical models for complex systems of chromophores interacting with various other degrees of freedom: while the chromophores interact with each other, so do the vibrational degrees of freedom of the protein scaffolding and the chromophores. This leads to a break-down of common approximations (Born-Oppenheimer, Condon, etc.) making simulations expensive and frustrating experiments. This is exacerbated by spectral congestion. Indeed, if the Frank-Condon approximation is assumed, oscillations in the pump-probe spectroscopy signal in the instantaneous-pulse limit correspond to coherent excitonic oscillations~\cite{Yuen-Zhou2012}, and a scheme to witness them using a series of pump-probe signals taken at several different pulse durations and extrapolating to the ultrashort-pulse limit has been proposed~\cite{Johnson2014}. To overcome these theoretical assumptions and practical difficulties experimental techniques have become ever more sophisticated by utilising multiple pulses, time delays, polarisations, and other properties of the incident light. However, even minor experimental issues such as a redshifted laser spectrum can mimic coherent excitonic oscillations in 2DES spectra~\cite{DeCamargo2017}, compounding the complication of identifying quantum coherences unambiguously.

Any worthwhile discussion on the quantum effects in EET must begin with an unambiguous quantifier of quantum coherence~\footnote{While it has been said that ``...coherence is theoretically well defined and also accessible to experimental quantification...''~\cite{Scholes:2017aa} the notion is by no means unique. Experimental accessibility then depends on which of these notions of coherence we seek to measure.} that is accessible experimentally. The size of the chromophores (a few nm) forbids any spatial resolution using far-field optical methods where the wavelengths are hundreds of nanometres. Notions of spatial quantum correlations are therefore not extractable~\footnote{Except perhaps with coherent 2D photoemission electron microscopy (PEEM) which provides 50-nm spatial resolution~\cite{Aeschlimann:2010aa}.}. Temporal quantum correlations are the alternative. One proposal suggested identifying the quantum behaviour of the vibrational motions that drive EET via phonon anti-bunching~\cite{OReilly:2014aa}. This however requires coherent ultrafast \emph{phonon} spectroscopy which remains challenging~\cite{Anderson:2018aa}. We therefore focus on a quantifier of quantum coherence that is experimentally accessible via non-linear ultrafast optical spectroscopy.

The notion of uniquely quantum correlations in time dates back to the work of Leggett and Garg, and their proposal for testing macroscopic realism (a certain classical limit of quantum mechanics) using inequality conditions~\cite{Leggett:1985aa,Emary:2013aa}. 
Intuitively, if a system undergoes a classical (possibly correlated) dynamical process, then interrupting with a non-invasive measurement and restarting it in the measured state would result in the same final state as if it had not been interrupted at all. On the contrary, interrupting a system undergoing a quantum dynamical process to measure and restart it may result in a different final state. It is this difference between quantum and classical states that quantum coherence measures endeavour to capture~\cite{Baumgratz:2014aa,Marvian:2016aa}. Mathematically, they capture the off-diagonal elements in the density matrix as expressed in a preferred basis: in the context of NSIT, this basis is determined by the interrupting operation. For a density matrix $\rho$, one such measure is
\begin{equation}
R(\rho) = \frac{1}{2}\left|\left|\rho-\Gamma(\rho)\right|\right|_{\text{tr}},
\end{equation}
where $\Gamma$ is the `non-invasive measurement' interruption that excises all off-diagonal elements in $\rho$ and $||X||_{\text{tr}}$ is the trace norm defined as the sum of the singular values of $X$. Evaluating $R(\rho)$ exactly requires the tomography of the quantum state $\rho.$ While possible in principle, it is expensive and possibly non-unique in practice. We therefore seek a lower bound on $R(\rho)$ that serves as an unambiguous witness to quantum coherence. Our witness is based on recent sharpenings and simplifications of the notion of Leggett and Garg, culminating in the so-called No-Signalling-In-Time (NSIT) witness~\cite{Kofler:2013aa,Robens:2015aa,Clemente:2016aa,Knee:2016aa}. This witness to temporal quantum coherence has also been measured in well-isolated engineered condensed matter quantum systems~\cite{Knee:2016aa}.

In this work we seek to measure the NSIT witness - a lower bound on $R(\rho),$ using ultrafast pump-probe spectroscopy. Pump-probe spectroscopy can be described as induced absorption and emission events from a nonstationary state~\cite{Abramavicius2009}. The matter Hamiltonian determines the natural basis for EET as the light-matter coupling is much weaker. Non-zero off-diagonal elements are evidence against a classical hopping in this basis. For weakly coupled chromophores, this is the regime of F\"roster transfer. In principle therefore, a non-zero $R(\rho)$ is evidence against models of EET mediated solely by the hopping of excitons, and in favour of quantum coherence playing a role. The same is true when the system under consideration is not isolated from its environment - as is the case for electronic excitations of chromophores in EET in relation to its vibrational modes, provided the Born approximation is satisfied at the time of this interruption. If not, as is likely in practice, subtleties arise~\cite{Knee:2018aa}.

%A considerable amount of effort has recently gone into modelling and understanding these spectroscopic techniques, especially 2DES and and four-wave mixing [...]. However, theoretical models can also be used in a different way. By understanding the spectroscopic methods from a theoretical point of view, experimental protocols can be designed that allow for a minimisation of experimental parameters while maximising the information extracted. Simpler yet, the derivation and use of witnesses allows for the answering of fundamental questions without delving too deep into quantitative measurements: why spend a considerable amount of time on measuring something quantitatively if we can show in a much simpler way that it is not there to begin with?

%We have, in previous work, laid the foundations for such witness for quantum coherences in coupled system-bath models. The presence of an environment (for instance vibrational modes of the chromophores or a protein scaffolding) introduces the possibilities of quantum coherences within the environment or between the system and the environment. However, in EET we are primarily interested in coherences within the system and a witness reporting on these alone has to be designed carefully. In this paper we will present the details of such a witness applied to EET and examine the conceptually easiest spectroscopic implementation.

The paper is organised as follows. In Sec. 2, we begin by adapting our previous work on measuring the No-Signalling-In-Time (NSIT) witness in a system coupled to a bath to suit the discussion of EET, showing that its non-zero value confirms the presence of quantum coherence. We show in Sec. \ref{sec:MDS} that for a model dimer system, all quantities needed for evaluating the NSIT witness are readily available from pump-probe spectroscopy. In fact, it can be obtained from very limited quantum process tomography of EET - needing only the population-to-population elements at three different times. We show that these elements can be extracted from a set of four two-pulse pump-probe experiments in Sec~\ref{sec:pps}. 
In Sec.~\ref{sec:Num}, we simulate our protocol numerically on allophycocyanin (APC), a molecular system studied in some detail for EET~\cite{Womick2009}. Our numerical results differ from the theoretical expectations, which we ascribe to a trade-off in the choice of pulse parameters. 
We conclude with a summary in Sec.~\ref{sec:conc} and an outlook on possible future avenues to spectroscopically measure the NSIT witness while limiting the experimental effort.

\section{The No-Signalling-In-Time Witness}

Previous work has considered the concept of quantum witnesses in systems coupled to baths~\cite{Li:2012aa,Knee:2018aa}. We derived three distinct witnesses of quantum coherence~\cite{Knee:2018aa} containing information about the presence of coherences between or within the system and bath. In this paper we focus on the second of these witnesses, labelled $W^b$, which provides information about the presence of coherences in time between system states. Mathematically,
\begin{equation}
R(\rho(t)) \geq 2W^b(t;T),
\end{equation}
where $T$ is the total time of the experiment. A positive value of $W^b(t;T)$ is thus an unambiguous witness of quantum coherence in EET. We briefly revisit the idea behind this witness before presenting an experimental scheme for obtaining it. For a more detailed discussion of $W^b(t;T)$ we refer to our previous work~\cite{Knee:2018aa}.

\begin{figure}
\centering
\includegraphics[width=1\linewidth]{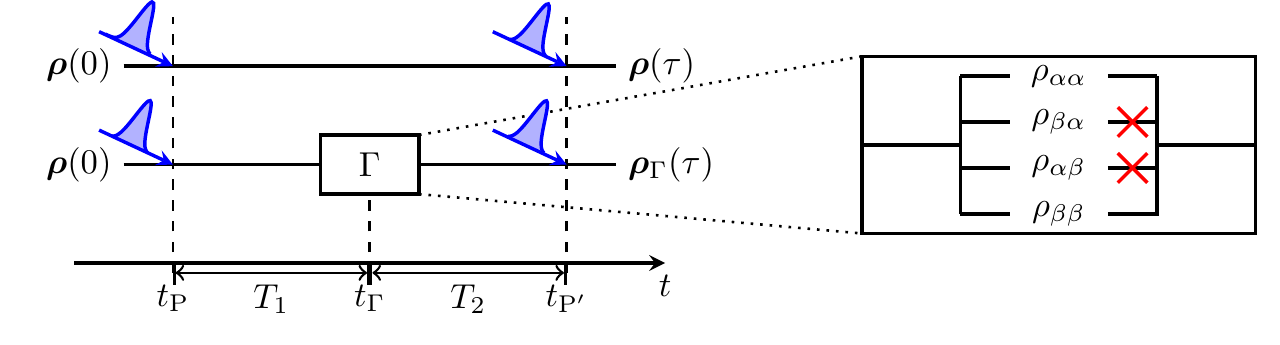}
\caption{Schematic depicting our proposed spectroscopic protocol. Two identical samples in identical states are exposed to the same pulses at the same time, with one being interrupted by the operation $\Gamma$. This operation is instantaneous and amounts to quantum population tomography and re-preparation without the off-diagonal elements (quantum coherence elements). The states are compared at the end.}
\label{Fig:1}
\end{figure}

The quantum state of a system coupled to a bath subject to a joint unitary evolution propagates between two times $t_{\text{P}}$ and $t_{\text{P}'}$ as (see Fig. \ref{Fig:1}),
\begin{equation}
\rho_{\text{SB}}(t_{\text{P}'})= U(t_{\text{P}'},t_{\text{P}}) {\boldsymbol \rho}_{\text{SB}}(t_{\text{P}}) U^{\dagger}(t_{\text{P}'},t_{\text{P}}) = \mathcal{U}_{t_{\text{P}'},t_{\text{P}}}[\rho_{\text{SB}}(t_{\text{P}})].
\label{Eq:U}
\end{equation}
In this work we will adopt the Born approximation for all $t_{\Gamma} \in (t_{\text{P}'},t_{\text{P}})$. This assumption is sufficient to allow the evolution to be partitioned at time $t_{\Gamma}$ and we may write
\begin{equation}
\rho_{\text{SB}}(t_{\text{P}'})= \mathcal{U}_{t_{\text{P}'},t_{\Gamma}}\circ\mathcal{U}_{t_{\Gamma},t_{\text{P}}}[\rho_{\text{SB}}(t_{\text{p}})].
\label{Eq:U2}
\end{equation}
%If the state at $t_{\Gamma}$ is not affected by some intervention at that time, then Eqs. (\ref{Eq:U}) and (\ref{Eq:U2}) will be the same. 
If, however, some operation $\Gamma$ is performed at time $t_{\Gamma}$ then the state,
\begin{equation}
\rho_{\text{SB}}^{\Gamma}(t_{\text{P}'}) =  \mathcal{U}_{t_{\text{P}'},t_{\Gamma}}\circ\Gamma\circ\mathcal{U}_{t_{\Gamma},t_{\text{P}}}[\rho_{\text{SB}}(t_{\text{p}})],
\end{equation}
may differ from the state in Eq. (\ref{Eq:U2}).
If a measurement of an observable $M$ is made on the system at time $t_{\text{P}'},$ the outcomes may be different depending on whether or not ${\Gamma}$ had affected its subsequent dynamics. The difference 
\begin{equation}
W^b(t_{\Gamma};\tau)  = \tr \left\{\left(M\otimes {\bf I}_{\text{B}}\right)\rho_{\text{SB}}(t_{\text{P}'})\right\} - \tr \{\left(M\otimes {\bf I}_{\text{B}}\right)\rho_{\text{SB}}^{\Gamma}(t_{\text{P}'})\},
\end{equation} 
where $\tau = t_{\text{P}'}-t_{\text{P}}$, can therefore serve as a witness to the quantum coherence of the state at time $t_{\Gamma}$ provided $\Gamma$ is \emph{non-invasive} in that it excises the off-diagonal elements in $\rho_{\text{SB}}(t_{\Gamma})$ instantaneously without affecting the subsequent dynamics in any way~\cite{Leggett:1985aa,Knee:2018aa}.

Since any joint system-bath quantum state at any time $t$ can be expressed as
\begin{equation}
 \rho_{\text{SB}}(t) =  \rho_{\text{S}}(t) \otimes \rho_{\text{B}}(t) + \gamma_{\text{SB}}(t),
\label{eq:BA}
\end{equation}
where $\gamma_{\text{SB}}(t)$ is the correlation between the system and the bath,
\begin{equation}
\begin{split}
W^b(t_{\Gamma};\tau) &= \left|\tr_{\text{S}}\left\{M\tr_{\text{B}}\left[\mathcal{U}_{t_{\text{P}'},t_{\Gamma}}\gamma_{\text{SB}}(t_{\Gamma})\right]\right\}\right.\\
& \left.+ \tr_{\text{S}}\left\{\mathcal{E}(M)\rho_{\text{S}}(t_{\Gamma})\right\} - \tr_{\text{S}}\left\{\mathcal{E}^{\Gamma}(M)\rho_{\text{S}}(t_{\Gamma})\right\}\right|,
\end{split}
\label{Eq:Wb}
\end{equation}
where $\mathcal{E}$ is a map containing the dynamics of the coupled system-bath evolution and the measurement performed on the system. $\mathcal{E}^{\Gamma}$ is defined in the same manner, with the operation $\Gamma$ preceding the natural evolution at $t_\Gamma$. The Born approximation has two effects on our witness: firstly, it forces $\gamma_{\text{SB}}(t) = 0\ \forall\ t$ causing the first term to vanish, and secondly the underlying evolution and measurement mappings, $\mathcal{E}(M)$ and $\mathcal{E}^{\Gamma}(M)$ are identical with the exception of the operation $\Gamma$ at $t_{\Gamma}$ for the interrupted case. As a result, $W^b(t_{\Gamma};\tau)$ reports exclusively on coherences between system eigenstates in the basis determined by $\Gamma$. 

Note that our witness $W^b(t_{\Gamma};\tau)$ is a two-time expectation value - the simplest instance of a multi-time response function~\cite{Abramavicius2009}, and potentially more informative than conventional 1D transient-absorption and transient-grating techniques. Our witness thus incorporates at least one crucial aspect of multidimensional spectroscopy that has been central in suggesting quantum coherence in EET in light-harvesting complexes~\cite{Engel:2007aa}.
For brevity, we suppress the time arguments of $W^b$ in the following.

%\AD{The excited states which are preparable and measurable in spectroscopy are determined by the relationship of the light pulses with the matter. In order to implement $\Gamma$, we may choose to measure and (re)prepare in an orthogonal basis of excited states: here we focus on the energy eigenbasis of the system (i.e. the exciton basis) and look to achieve this by tuning light pulses to exclusively target one transition or the other.\\As such, $W^b$ gives the difference in the populations of one eigenstate of the system when interrupted at time $t_{\Gamma}$. }

To proceed, the quantum process formalism is particularly useful as it allows us to relate any reduced density matrix of the system at time $t_{\text{P}'}$ to its state at time $t_{\text{P}}$ via
\begin{equation}
\boldsymbol{\rho}_{\text{S}}(t_{\text{P}'}) =\boldsymbol{\chi}(t_{\text{P}'}-t_{\text{P}})\boldsymbol{\rho}_{\text{S}}(t_{\text{P}}).
\end{equation}
$\chi(t)$ is a degree-4 tensor called the process tensor~\cite{Yuen-Zhou:2011aa}. Thus any element of the density matrix at any time $t>t_{\text{P}}$ is
\begin{equation}
\rho_{ij}(t) = \sum_{pq} \chi_{ijpq}(t-t_{\text{P}})\rho_{pq}(t_{\text{P}}).
\end{equation}
This allows us to express the unitary evolution in terms of the process tensor and express
\begin{equation}
\tr_{\text{S}}\left\{\mathcal{E}(M)\rho_{\text{S}}(t_{\Gamma})\right\} = \sum_{pq} \chi_{iipq}(t_{\text{P}'}-t_{\Gamma})\rho_{pq}(t_{\Gamma}),
\end{equation}
\begin{equation}
\tr_{\text{S}}\left\{\mathcal{E}^{\Gamma}(M)\rho_{\text{S}}(t_{\Gamma})\right\} = \sum_{pq} \chi_{iipq}(t_{\text{P}'}-t_{\Gamma})\Gamma\left[\rho_{pq}(t_{\Gamma})\right],
\end{equation}
where $M = \left|i\right\rangle\left\langle i\right|$ measures the population in the eigenstate $\left|i\right\rangle$. Since the state is initially prepared at time $t_{\text{P}}$ then we can further expand the final quantities in the above equations to obtain
\begin{equation}
\tr_{\text{S}}\left\{\mathcal{E}(M)\rho_{\text{S}}(t_{\Gamma})\right\} = \sum_{pqrs} \chi_{iipq}(t_{\text{P}'}-t_{\Gamma})\chi_{pqrs}(t_{\Gamma}-t_{\text{P}})\rho_{rs}(t_{\text{P}}),
\label{Eq:E1}
\end{equation}
\begin{equation}
\tr_{\text{S}}\left\{\mathcal{E}^{\Gamma}(M)\rho_{\text{S}}(t_{\Gamma})\right\} = \sum_{pqrs} \chi_{iipq}(t_{\text{P}'}-t_{\Gamma})\Gamma\left[\chi_{pqrs}(t_{\Gamma}-t_{\text{P}})\rho_{rs}(t_{\text{P}})\right].
\label{Eq:G2}
\end{equation}
Choosing the operation $\Gamma$ such that it eliminates all coherences in the system at time $t_{\Gamma}$ then alters Eq. (\ref{Eq:G2}) to give
\begin{equation}
\tr_{\text{S}}\left\{\mathcal{E}^{\Gamma}(M)\rho_{\text{S}}(t_{\Gamma})\right\} = \sum_{prs} \chi_{iipp}(t_{\text{P}'}-t_{\Gamma})\chi_{pprs}(t_{\Gamma}-t_{\text{P}})\rho_{rs}(t_{\text{P}}).
\label{Eq:G3}
\end{equation}
As we have assumed the Born approximation, we can contract the two process tensors in Eq. (\ref{Eq:E1}) into one for the entire time. This cannot be done in Eq. (\ref{Eq:G3}) as the summation is incomplete. This is the mathematical essence of our witness.

Combining Eqs. (\ref{Eq:E1}) and (\ref{Eq:G3}) into Eq.~(\ref{Eq:Wb}) and invoking the Born approximation, we have
\begin{widetext}
\begin{equation}
W^b =\left| \sum_{rs} \left\{\chi_{iirs}(t_{\text{P}'}-t_{\text{P}}) - \sum_p \chi_{iipp}(t_{\text{P}'}-t_{\Gamma})\chi_{pprs}(t_{\Gamma}-t_{\text{P}})\right\}\rho_{rs}(t_{\text{P}})\right|.
\label{Eq:Wb2}
\end{equation}
\end{widetext}
Note that this quantity depends on the state $\left|i\right\rangle$, the projection onto which we chose to be our final measurement. Intuitively, this definition captures the following: the first term within the summation captures all dynamics from $\left|r\right\rangle\left\langle s\right|$ at $t_{\text{P}}$ to $\left|i\right\rangle\left\langle i\right|$ at $t_{\text{P}'}$ via all possible routes, while the second term excludes any dynamics between these arising from coherences in the exciton basis at some arbitrarily chosen intermediate time $t_{\Gamma}$. If these two terms are equal, then there is no dynamics via states involving coherence in the exciton basis at $t_{\Gamma}$ indicating that the dynamics is, indeed, incoherent. Any non-zero value for $W^b$ then indicates that quantum coherences are indeed present at time $t_{\Gamma}$. We can simplify the witness further by starting, for instance, with $\rho(t_{\text{P}}) = \left|\alpha\right\rangle\left\langle\alpha\right|$. This will become useful below when we adapt the witness to suit a pump-probe experiment for a dimer system.

\section{Model Dimer System}
\label{sec:MDS}

We now present an explicit expression for $W^b$ for a model dimer system and derive a spectroscopic protocol which is, in principle, capable providing the necessary elements of the process tensor without the need of full quantum process tomography (QPT) thereby reducing the experimental (and computational) cost significantly.

\subsection{System-Bath Hamiltonian}
We model the excited state dynamics of a dimer coupled to a bath of harmonic oscillators, one per site. The corresponding Hamiltonian takes the form of a Frenkel-Holstein Hamiltonian,
\begin{equation}
\hat{H}_{\text{S}} = \sum_{i=\left\{ \text{a,b}\right\}} \varepsilon_i \hat{a}_i^{\dagger}\hat{a}_i + J\left(\hat{a}_{\text{a}}^{\dagger}\hat{a}_{\text{b}} + \hat{a}_{\text{b}}^{\dagger}\hat{a}_{\text{a}}\right)
\label{Eq:HS}
\end{equation}
\begin{equation}
\hat{H}_{\text{B}} = \hbar \sum_{i=\left\{\text{a,b}\right\}} \omega_i \left(\hat{b}_i^{\dagger}\hat{b}_i + \frac{1}{2}\right) 
\end{equation}
\begin{equation}
\hat{H}_{\text{SB}} = -\hbar\sum_{i=\left\{\text{a,b}\right\}} \omega_ig_i \hat{a}_i^{\dagger}\hat{a}_i\left(\hat{b}_i^{\dagger} + \hat{b}_i\right),
\label{Eq:HSB}
\end{equation}
where a and b are the site indices and $\hat{a}_i^{\dagger}/\hat{a}_i$ and $\hat{b}_i^{\dagger}/\hat{b}_i$ are exciton and phonon creation/annihilation operators respectively. The dimensionless exciton-phonon coupling parameter, $g_i$, is related to the Huang-Rhys parameter, $S_i$, as $S_i = g_i^2/2$. For a more detailed discussion of the Frenkel-Holstein model see the work by Barford et al.~\cite{Barford:2014aa,Marcus:2014ab,Barford:2017aa}.

\begin{figure}
\centering
\includegraphics[width=1\linewidth]{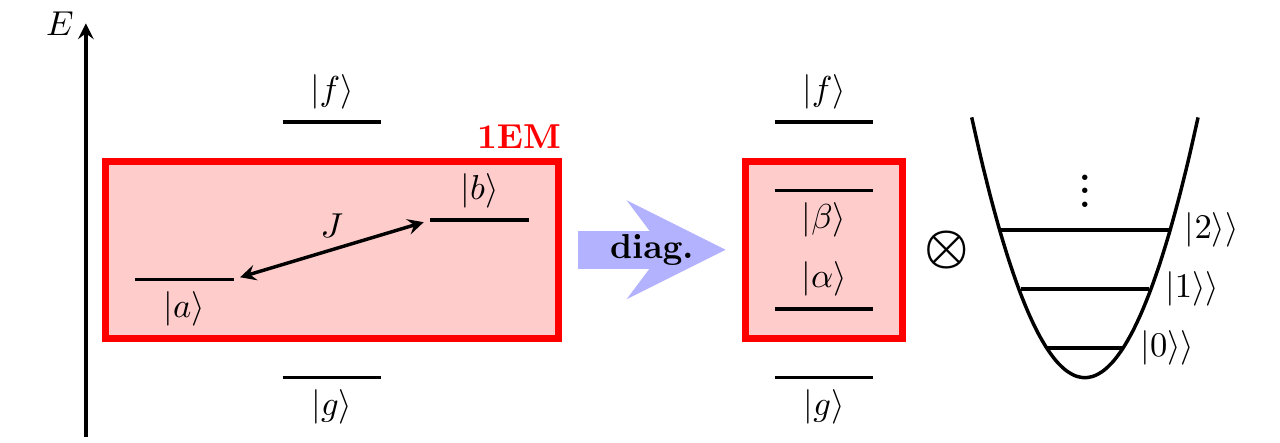}
\caption{Schematic depiction of the basis set used. The site basis (left) is diagonalised and expanded by local harmonic oscillators. The excitonic and phononic degrees of freedom are then coupled via $\hat{H}_{\text{SB}}$. The EET dynamics occur in the singly excited manifold (1EM).}
\label{Fig:2}
\end{figure}

The basis states of the uncoupled Hamiltonian, as shown in Fig. \ref{Fig:2} (left), are the electronic states, $\left\{\left|g\right\rangle,\left|a\right\rangle,\left|b\right\rangle,\left|f\right\rangle\right\}$ and the ladder of localised phonon states of the harmonic oscillator, $\left\{\left.\left|n\right\rangle\right\rangle_m\right\}$ of $n$ phonons on site $m$. We construct a basis by diagonalizing the uncoupled Hamiltonian, resulting in product states linking one electronic eigenstate to a ladder of delocalised phonons. The electronic eigenstates are the electronic ground state, $\left|g\right\rangle$, the two states forming the singly excited state manifold (1EM), $\left|\alpha\right\rangle$ and $\left|\beta\right\rangle$, and the state forming the doubly excited state manifold (2EM), $\left|f\right\rangle$. $\left|\alpha\right\rangle$ and $\left|\beta\right\rangle$ correspond to the bonding and anti-bonding orbitals of a dimer system if $J<0$.

We are interested in the excited state dynamics occurring in the singly-excited manifold (1EM) of the system under consideration. Experimentally, non-linear spectroscopic techniques are commonly used to elucidate these dynamics and to determine the existence of quantum coherences between states. Most prominently, two-dimensional electronic spectroscopy (2DES) is used to detect off-diagonal peaks in the spectra, oscillations of which are commonly interpreted as signatures for quantum coherences~\cite{Engel:2007aa}. 
Note that, while such oscillations are consistent with a quantum coherent model, they could also be consistent with certain classical models. Experimentally simpler is pump-probe spectroscopy in which the sample is exposed to two short pulses of light with a time delay, $\tau$, where the first pulse (pump) generates an excited state in the sample which is then measured by the second pulse (probe). We use this spectroscopic set-up to develop a protocol making our previous theoretical findings experimentally accessible. Somewhat counterintuitively, our protocol allows quantum coherence to be revealed (if it exists) without the need for more complicated types of spectroscopy, such as 2DES. This is because quantum coherence is not probed directly, but only indirectly through its implicit deletion under the $\Gamma$ operation, which we aim to achieve using only two-pulse pump-probe experiments. Moreover, our protocol allows for classical models (that have zero quantum coherence) to be decisively ruled out.

\subsection{Quantum Coherence Witness}
\label{sec:wit}

Full quantum process tomography is practically very expensive as the number of unknowns scales as $d^4 - d^2$, where $d$ is the dimensionality of the Hilbert space of the system. This is even before problems of non-uniqueness arise in dealing with realistic scenarios~\cite{Ruckebusch2012,Gururangan2019}. In its current form, our witness in Eq. (\ref{Eq:Wb2}) requires far less than full knowledge of the process tensor, albeit for two distinct evolution periods. However, some additional assumptions can further reduce the workload significantly: (i) the most drastic assumption (and most difficult one to realise in experiment) is that at $t_{\text{P}}$ the system is in the state $\left|\alpha\right\rangle$. Then Eq. (\ref{Eq:Wb2}) reduces to,
\begin{equation}
W^b =\left| \chi_{ii\alpha\alpha}(t_{\text{P}'}-t_{\text{P}}) - \sum_p \chi_{iipp}(t_{\text{P}'}-t_{\Gamma})\chi_{pp\alpha\alpha}(t_{\Gamma}-t_{\text{P}})\right|,
\end{equation}
as $\rho_{\alpha\alpha}(t_{\text{P}}) = 1$. A similar assumption is required to enable us to achieve $\Gamma$: namely, the ability to selectively prepare and measure $\left|\alpha\right\rangle$ and $\left|\beta\right\rangle$. (ii) trace preservation, that is no excitonic dissipation, implies~\cite{Yuen-Zhou:2014ab}
\begin{equation}
\sum_{i} \chi_{iipq}(t) = \delta_{pq}\quad \forall p,q.
\end{equation}
For a dimer system with eigenstates $\left|\alpha\right\rangle, \left|\beta\right\rangle,$ trace preservation and setting $i = \alpha$ leads to 
\begin{equation}
\begin{split}
W^b &=\left| \chi_{\alpha\alpha\alpha\alpha}(\tau) + \chi_{\alpha\alpha\alpha\alpha}(T_1)\left(1-\chi_{\alpha\alpha\alpha\alpha}(T_2)\right)\right.\\
&\qquad\left. + \left(1-\chi_{\alpha\alpha\alpha\alpha}(T_1)\right)\chi_{\beta\beta\beta\beta}(T_2) - 1\right|,
\end{split}
\label{Eq:Wb22}
\end{equation}
where $T_1 = t_{\Gamma} - t_{\text{P}}$, $T_2 = t_{\text{P}'} - t_{\Gamma}$, and $\tau = T_1 + T_2 =  t_{\text{P}'}-t_{\text{P}}$. The significance of Eq. (\ref{Eq:Wb22}) is that it allows us to witness quantum coherence in the system by merely determining two - indeed only the population-to-population elements of the quantum process tensor at three distinct times. This is a significant reduction of the workload compared to full quantum process tomography and one of our main results.

\section{Pump-probe spectroscopy}
\label{sec:pps}

We now derive a spectroscopic protocol which, in principle, witnesses quantum coherences experimentally and numerically simulate it in Sec.~\ref{sec:Num} by integrating the time-dependent Schr\"{o}dinger equation for the time-dependent semi-classical light-matter Hamiltonian.  
To benchmark these simulations, we theoretically calculate $W^b$ via the process tensor for the excited state dynamics, governed by the sum of the Hamiltonians in Eqs. (\ref{Eq:HS}) - (\ref{Eq:HSB}). This essentially replaces the semi-classical light-matter interaction with perfect preparation and measurement of the 1EM states $\left|\alpha\right\rangle, \left|\beta\right\rangle$ that we wish to target.

\subsection{Excited State Dynamics}
\label{sec:ExSD}

The model dimer introduced in Sec. 3.1 is subjected to a time-dependent classical field given by
%\begin{equation}
%\begin{split}
\begin{eqnarray}
\label{Eq:Pulset}
\hat{V}(t) &=& -\sum_n\sum_{i} \hat{\boldsymbol{\mu}}_{i}\cdot{\bf E}_n(t)\\
&=& -\sum_{ni} \boldsymbol{\hat{\mu}}_{i}\cdot{\bf e}_n\left(\hat{a}_i^{\dagger}+\hat{a}_i\right)e^{-\frac{(t-t_n)^2}{2\sigma_n^2}}\left[\text{e}^{-\text{i}\omega_n t} + \text{e}^{\text{i}\omega_n t}\right], \nonumber
\end{eqnarray}
%\end{split}
%\end{equation}
where $i$ denotes the excitonic states, $n$ the light pulses of frequency $\omega_n,$ $t_n$ the central time of the Gaussian envelope of these pulses, $\sigma_n$ their width,   and ${\bf e}_n$ their polarisation. The intensity of the field is denoted by $\eta_n.$ When interacting with the system the oscillating light field will have one term significantly closer to resonance than the other. While computationally not less expensive, making the Rotating Wave Approximation (RWA), which neglects the off-resonant oscillation, makes the theoretical calculation of transition probability amplitudes easier. We use the RWA throughout this work.

\begin{figure}[h]
\centering
\includegraphics[width=1\linewidth]{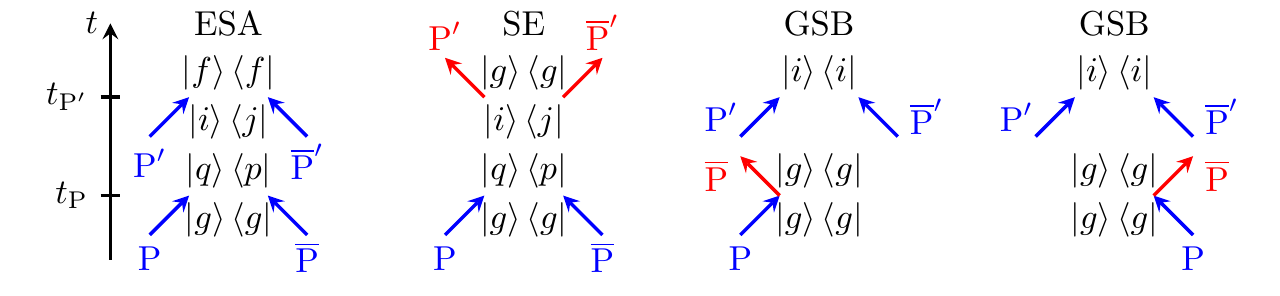}
\caption{First-order non-linear optoelectronic processes in a general system when interacting with two pulses (P and $\text{P}'$) at two distinct times shown as double-sided Feynman Diagrams. ESA: Excited State Absorption; SE: Stimulated Emission; GSB: Ground-State Bleach. Each process involves the 1EM.}
\label{Fig:3}
\end{figure}

To simulate spectroscopic experiments we initialise the system and environment in their respective ground states, in effect leading to simulations of experiments at 0 K. The extension to finite temperatures is straightforward. The interaction of two subsequent pulses with a sample initialised in this way will give rise to a multitude of processes between the ground and excited states.  To first-order these are either of Excited State Absorption (ESA), Stimulated Emission (SE), or Ground-State Bleach (GSB), as illustrated in Fig. \ref{Fig:3}. Each of these processes will contribute to the overall signal observed in a pump-probe experiment. Following Yuen-Zhou et al.~\cite{Yuen-Zhou:2014ab}, we calculate these contributions by considering the transition probability amplitudes
\begin{equation}
\Omega_{ij}^p = \tilde{E}_p(\omega_{ij})\boldsymbol{\mu}_{ij}\cdot{\bf e}_p,
\label{Eq:Omega}
\end{equation}
where $\tilde{E}_p(\omega_{ij})$ is the Fourier-transform of the Gaussian pulse in Eq.~(\ref{Eq:Pulset}) and $\boldsymbol{\mu}_{ij}$ is the transition dipole moment between states $\left|j\right\rangle$ to $\left|i\right\rangle.$ The dynamics occurring within the system between the two pulses are quantified by the process tensor describing the natural evolution. Each contribution (ESA/SE/GSB) can then be expressed as the product of three different factors summed over each possible combination of states: 
	(1). A factor for exciting the true ground state with a given pulse (P) into a given excited `state'; 
	(2). A factor representing transition from that `state' into another `state' within the 1EM under natural evolution; and 
	(3). A factor of reaching a given target `state' under interaction with the second pulse (P$'$).

	For instance, the ESA contributions are given by
\begin{equation}
S_{\text{ESA}}(\tau) = \sum_{ijpq} \Omega_{qg}^{\text{P}}\Omega_{gp}^{\bar{\text{P}}}  \Omega_{fi}^{\text{P}'}\Omega_{jf}^{\bar{\text{P}}'}\chi_{ijqp}(\tau),
\label{Eq:ESA}
\end{equation}
where the factors have been expressed using the transition probability amplitudes. Note that the `states' are sometimes non-Hermitian operators, and their appearance is an idiosyncrasy of perturbation theory.
Similar expressions for the SE and GSB are given by
\begin{equation}
S_{\text{SE}}(\tau) = -\sum_{ijpq}\Omega_{gi}^{\overline{\text{P}}'}\Omega_{qg}^{\text{P}}\Omega_{gp}^{\overline{\text{P}}}\Omega_{jq}^{\text{P}'}\chi_{ijqp}(\tau),
\label{Eq:SE}
\end{equation}
\begin{equation}
S_{\text{GSB}} = -\sum_{ip}\Omega_{ig}^{\text{P}'}\Omega_{gp}^{\overline{\text{P}}}\Omega_{pg}^{\text{P}}\Omega_{gi}^{\overline{\text{P}}'},
\label{Eq:GSB}
\end{equation}
where $i,j,p,q$ label states within the 1EM. The total signal of a general pump-probe experiment is then given as the sum of these contributions and will, in general, be a function of all elements of the process tensor, 16 for a two-level 1EM. 

In order to be experimentally (and computationally) cheaper than QPT the number of these unknowns can be reduced by making judicious choices of experimental parameters, which will directly inform the simulation of the resulting experiments. Furthermore, in order for our quantum witness $W^b$ to inform on the presence of coherences within the natural evolution we have to explicitly exclude the possibilities of creating such coherences with the light pulses. This is because the witness only informs on the presence of any coherence, without regard as to where they have arisen from. We therefore require the initially prepared state to be an eigenstate, and $p = q$ in Eqs.~(\ref{Eq:ESA})-(\ref{Eq:GSB}). This constraint eliminates eight elements in the process tensor. By tailoring the light field to only prepare and probe excitonic eigenstates we can eliminate a further four contributions, reducing the equations above to four unknown elements, namely the population-to-population elements, $\chi_{iipp}$. The equations then reduce to,

%These contributions can be simplified by making some (justified) approximations to ease the computational and - subsequently - experimental effort. First, we will ignore the creation of coherences when the matter interacts with the light by choosing the pulses in such a way that they excite one excited state selectively. This choice of pulses has been labelled 'Maximally Discriminant Choice' (MDC) and it will form a central assumption in this work. There are some issues concerning MDC which we will discuss below.

%By ignoring the creation of coherences by interaction with light we can restrict the labels for the 1EM in Eqs. (\ref{Eq:ESA})-(\ref{Eq:GSB}) to $p=q$ and $i=j$. By then also making use of $\Omega_{rs}^{p} = \left[\Omega_{sr}^{\overline{p}}\right]^*$ and defining $\Pi_{rs}^p = \left|\Omega_{rs}^p\right|^2$ as the transition probability we can re-write the equations above as,
\begin{equation}
\mathcal{S}_{\text{ESA}}(\tau) = \sum_{pq} \Pi_{fq}^{\text{P}'}\Pi_{pg}^{\text{P}}\chi_{qqpp}(\tau),
\label{Eq:ESA2}
\end{equation}
\begin{equation}
\mathcal{S}_{\text{SE}}(\tau) = -\sum_{pq} \Pi_{gq}^{\text{P}'}\Pi_{pg}^{\text{P}}\chi_{qqpp}(\tau),
\label{Eq:SE2}
\end{equation}
\begin{equation}
\mathcal{S}_{\text{GSB}}(\tau) = -\sum_{pq}\Pi_{gq}^{\text{P}'}\Pi_{pg}^{\text{P}},
\label{Eq:GSB2}
\end{equation}
where for our model dimer $p,q\in\left\{\alpha,\beta\right\},$ $\Pi_{pg}^{\text{P}} = \Omega_{pg}^{\text{P}}\Omega_{gp}^{\overline{\text{P}}}$ and similarly for $\text{P}'$. To find the four elements of the process tensor we require four linearly-independent signals containing information about the excited state dynamics within our electronic dimer system. Indeed, the signal can be expressed as
\begin{equation}
\begin{split}
\mathcal{S}(\tau) &= \mathcal{S}_{\text{ESA}}(\tau) + \mathcal{S}_{\text{SE}}(\tau) + \mathcal{S}_{\text{GSB}}(\tau)\\
&= \sum_{pq} \left[\Pi_{fq}^{\text{P}'}\Pi_{pq}^{\text{P}} - \Pi_{gq}^{\text{P}'}\Pi_{pg}^{\text{P}}\right]\chi_{qqpp}(\tau) -\sum_{pq}\Pi_{gq}^{\text{P}'}\Pi_{pg}^{\text{P}}\\
&= \sum_{qp} M_{qp}(\text{P}',\text{P})\chi_{qqpp}(\tau)  - G(\text{P}',\text{P}).
\label{Eq:S1}
\end{split}
\end{equation}
Selecting four different pulse sequences $(\text{P}',\text{P})$ with the same delay time then allows us to re-write Eq. (\ref{Eq:S1}) in vectorised form as,
\begin{equation}
{\bf S}(\tau) = {\bf M}\boldsymbol{\vec{\chi}}(\tau) - {\bf G},
\label{Eq:Mat}
\end{equation}
where $\boldsymbol{\vec{\chi}}(\tau)$ is the vectorised reduced process tensor with the four entries $\chi_{qqpp}(\tau)$ for $p,q\in\left\{\alpha,\beta\right\}$. In order to recover the process tensor elements from this equation we require the matrix ${\bf M}$ to be non-singular and well-conditioned. As the elements of the matrix depend directly on the pulses used, their choice is crucial in order to a well-conditioned ${\bf M}$ matrix. 

\subsection{Choice of Pulse Sequences}
\label{sec:pulse}

\begin{figure}
\centering
\includegraphics[width=1\linewidth]{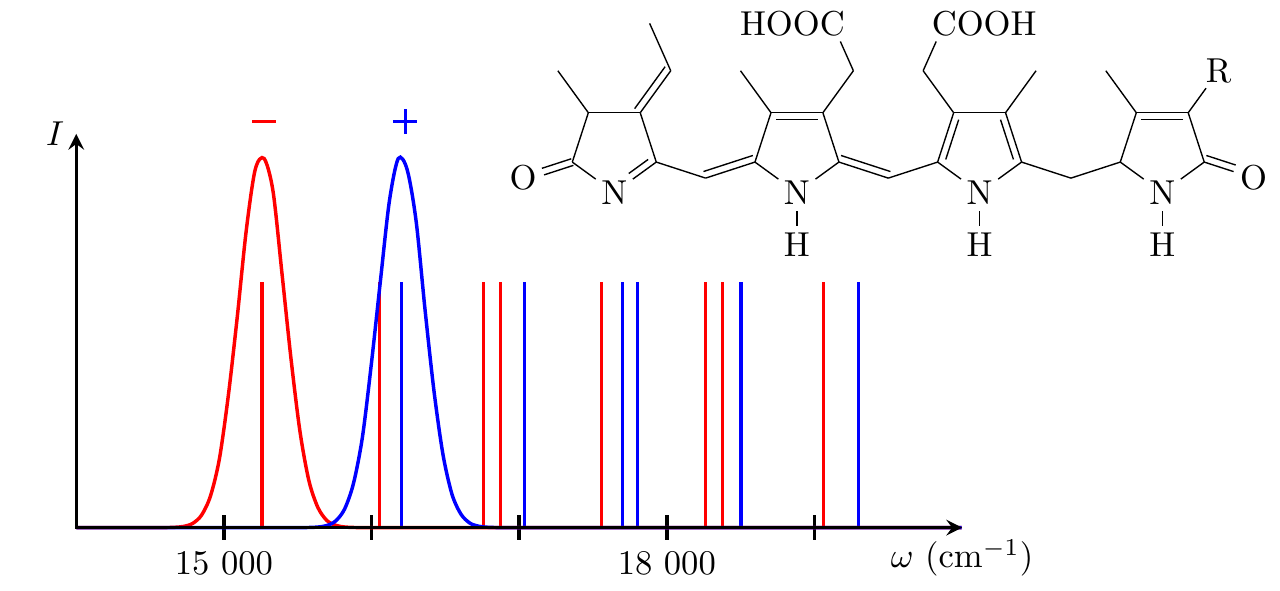}
\caption{Absorption energies of a allophycocyanin (APC) dimer~\cite{Womick2009} not coupled to the bath ($\hat{H}_{\text{SB}} = 0$) and structural formula of the chromophore phycoerythrobilin (R = C$_2$H$_3$).The pulses targeting the vibrationless excited states are indicated as well.}
\label{Fig:4}
\end{figure}

Eq. (\ref{Eq:Mat}) is derived on the assumption that we can create and measure an eigenstate of the system under investigation, i.e. pulses of light selectively interact with only one electronic transition. We therefore need to ensure that the pulses have large spectral overlap with one transition and vanishing overlap with all others. For a dimer with two electronic states and no environment the choice of pulses is straight-forward: by de-tuning the pulses in frequency from the two transitions in such a way that the energetic gap between the pulses is larger than between the transitions the cross-talk can be minimised. In such a way the transition probability amplitudes to coherent states vanish and the approximations made above in Sec.~\ref{sec:ExSD}, allowing for an accurate and precise recovery of the process tensor elements. For a system coupled to a bath both transitions for a dimer acquire a vibronic tail, which can overlap. Fig. \ref{Fig:4} presents the absorption spectra of the allophycocyanin (APC), the dimer we use for our simulations later. The relative spacings are typical of commonly encountered chemical species in the $\hat{H}_{\text{SB}} = 0$ limit. The two vibrational manifolds of the two electronic transitions are shown in red and blue.

In order to exclude the possibility of higher-order processes the incident light pulses need to be short such that excitations occur on a timescale much shorter than the natural evolution. In order to excite $|\alpha\rangle,|\beta\rangle$ exclusively, the pulses must be much narrower in frequency than the gap between the vibrational ladders associated with them. Simultaneous temporal and spectral resolution is however limited for Fourier-limited pulses. This directly impacts the assumption we made in Sec.~\ref{sec:ExSD} and the evaluation of our witness below.

%It is within these constraint of the Fourier transform that the pulses need to be chosen: ideally they are delta pulses in both, frequency and time.
%Clearly visible is that even in the case of vanishing coupling the tails overlap significantly. We can therefore not target electronic transitions without exciting coherences while being broadband with respect to the vibrational degrees of freedom. 
%
%This directly impacts the assumption we made above: by not being able to exclude the possibility of creating coherences with the light-pulse the witness will report on coherences generated by the light as well as the system dynamics, without distinguishing between the two. 
%
%This is the 'Fourier transform' of a problem commonly encountered in experimental spectroscopy: 

In the following, we choose the pulses resonant with the electronic transitions and labelled them `+' and `-' for the higher and lower energy transitions respectively, as illustrated in Fig.~\ref{Fig:4}. With this toolbox of two pulses we construct four distinct pump-probe experiments and describe how these can be used to measure the witness derived above.

\subsection{Spectroscopic Protocol}
\label{sec:spec}

\begin{figure}
\centering
\includegraphics[width=1\linewidth]{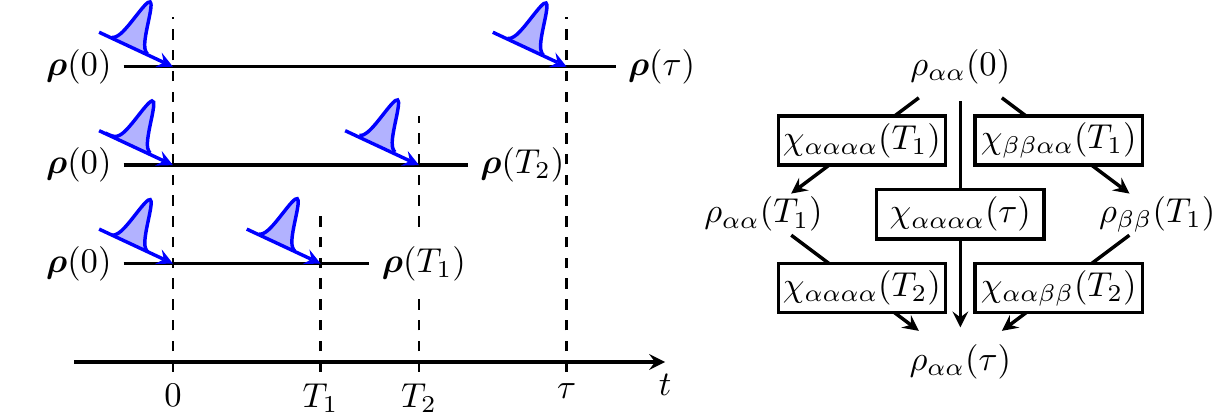}
\caption{Schematic showing our proposed experimental implementation of the protocol. Three identical examples are pumped with the same pulse and probed at different times to extract population-to-population elements of the process tensor. Comparing the result from two distinct times, $T_1$ and $T_2$, making up an overall time, $\tau$, allows a statement about the presence of quantum coherences at $T_1$.}
\label{Fig:5}
\end{figure}

As shown in Eq.~(\ref{Eq:Wb22}), the evaluation of $W^b$ via the implementation of $\Gamma$ can be replaced with partial quantum process tomography at three distinct delay times. In particular, it requires the population-to-population elements of the process tensor at times $T_1$, $T_2$, and $\tau = T_1 + T_2$.  
The experiment therefore breaks down into three distinct experiments per pulse sequence as illustrated in Fig. \ref{Fig:5}. 
Three identical samples are exposed to the same pump pulse and probed with the same pulse at the three delay times. 
The results of the four different pulse sequences allows, via Eq.~(\ref{Eq:Mat}), for the recovery of the population-to-population elements at different delay times which allows for the calculation of the witness. 

Note that the derivation of the witness $W^b$ in Eq.~(\ref{Eq:Wb22}) assumed the excision of all coherences within the systems at time $t_{\Gamma},$ as illustrated in Fig. \ref{Fig:1}. This may be possible in the future where pulses may be designed which destroy the off-diagonal terms, for instance by randomising their phases. In a previous measurement of the NSIT witness~\cite{Knee:2016aa} this was achieved by entangling the system of interest with an additional system. Whether similar or independent schemes for implementing $\Gamma$ can be devised for EET in light-harvesting complexes is beyond the scope of this work.
 %the design of such pulse(s) is outside of the scope of this work.}
We also note the work of Moreira and Semi\~{a}o~\cite{Moreira:2019aa}, where a different mathematical operation is proposed to test a distinct but related witness, but is not translated into an ostensibly feasible spectroscopic operation.

%Alternatively, these elements can be determined at different times and for each triple $\tau = T_1 + T_2$ the witness can then be calculated.

%As laid out in Sec. 3.2 we require the population-to-population elements of the process tensor for three distinct natural evolution times: $T_1$, $T_2$, and $\tau = T_1 + T_2$. According to Sec. 3.4 we can measure these elements using pump-probe spectroscopy if we choose the pulses judiciously (see Sec. 3.5). For a two-level system with two targeted pulses we can construct four pump-probe experiments that in principle fulfil these requirements. 

%Due to the discussion in Sec. 3.2, $\Gamma$ can be replaced with partial quantum process tomography at three distinct delay times, as laid out above. The experimental set up therefore breaks down into three distinct experiments per pulse sequence (see Fig. \ref{Fig:5}): three identical samples are exposed to the same pump-pulse and probed with the same pulse at the three delay times mentioned above. The results of the four different pulse sequences allows for the recovery of the population-to-population elements at different delay times which allows for the calculation of the witness. Alternatively, these elements can be determined at different times and for each triple $\tau = T_1 + T_2$ the witness can then be calculated.

\section{Numerical Results}
\label{sec:Num}

We now present numerical simulations of the spectroscopic protocol set out above on a dimer of phycoerythrobilin, a chromophore in APC~\cite{Womick2009} found in blue-green (cyanobacteria) and red algae.
The parameters used in our simulations are given in Tab. \ref{tbl:parameter}.
Details on our computational methodology is provided in App.~(\ref{App:Comp}). We briefly highlight the well-recognised computational complexity of simulating non-linear spectroscopy experiments in general and the advantages offered by our strategy of witnessing quantum coherence.

Firstly, the size of the Hilbert space in which the Hamiltonians in Eqs. (\ref{Eq:HS}) - (\ref{Eq:HSB}) reside scales unfavourably with the number of chromophores and phonons considered. For a system of $N_{\text{sites}}$ sites with localised harmonic oscillators with $N_{\text{phon}}$ vibrationally excited states the Hilbert space of all excitations has the dimensionality
\begin{equation}
d = N_{\text{sites}}^2\left(N_{\text{phon}} + 1\right)^{N_{\text{sites}}}.
\end{equation}
For instance, for a FMO complex with 7 sites and 9 phonons on each site (10 vibrational levels for a single harmonic oscillator), $d = 4.9 \times 10^{8}$, which would require simulating about 10$^{34}$ experiments for full QPT. Even for a dimer with 4 phonons per site, which has $d=100,$ this number would be 10$^{8}.$ 
Using our protocol, witnessing quantum coherence using $W^b$ requires simulating only 12 experiments.

Secondly, the need for averaging to simulate samples in realistic experiments multiples the cost due to the direct simulation of different orientations. We estimate this number to be in the thousands for our model dimer. The independence of the different orientations, however, make the simulation highly parallelisable. 

As shown in Sec.~(\ref{sec:ExSD}), the selectivity of pump and probe pulses is vital to our strategy of measuring our witness. To that end, we use two different methods. The first uses the polarisation of pulses and the relative orientation of the transition dipole moments in a single molecule experiment (Sec.~\ref{sec:single}). The second uses energetic targeting of the transitions and proceeds by averaging over an ensemble of dimers (Sec.~(\ref{sec:ens}).

%model system often used to research EET, namely a dimer of phycoerythrobilin, a chromophore found in allophycocyanin. We will use two different methods to achieve pulse selectivity: first by utilising the polarisation of pulses and the relative orientation of the transition dipole moments in a single dimer experiment, and secondly by averaging isotropically over an ensemble of dimers and using energetic targeting of the transitions. The values for the parameters in the Hamiltonians in Eqs. (\ref{Eq:HS}) - (\ref{Eq:HSB}) are given in Tab. \ref{tbl:parameter}.

\begin{table}[h]
\small
  \caption{\ System and bath parameters for a dimer of phycoerythrobilin~\cite{Womick2009} and light pulses used for simulations.}
  \label{tbl:parameter}
  \begin{tabular*}{0.48\textwidth}{@{\extracolsep{\fill}}ll}
    \hline
    Parameter (unit) & Value \\
    \hline
    $\varepsilon_a$ (cm$^{-1}$) & 15 300 \\
    $\varepsilon_b$ (cm$^{-1}$) & 16 200  \\
    $J$ (cm$^{-1}$) & -162 \\
    $\omega_a$ (cm$^{-1}$)  & 800 \\
    $\omega_b$ (cm$^{-1}$) & 1500 \\
    $g_a$ & 0.1 \\
    $g_b$ & 0.15 \\
    $\sigma_p$ (cm$^{-1}$/fs) & 322/103 \\
    $\eta_p$ (eV ps/D) & $5\times 10^{-6}$ \\
    \hline
  \end{tabular*}
\end{table}

\subsection{Single Dimer Spectroscopy}
\label{sec:single}

\begin{figure*}
\centering
\includegraphics[width=0.8\linewidth]{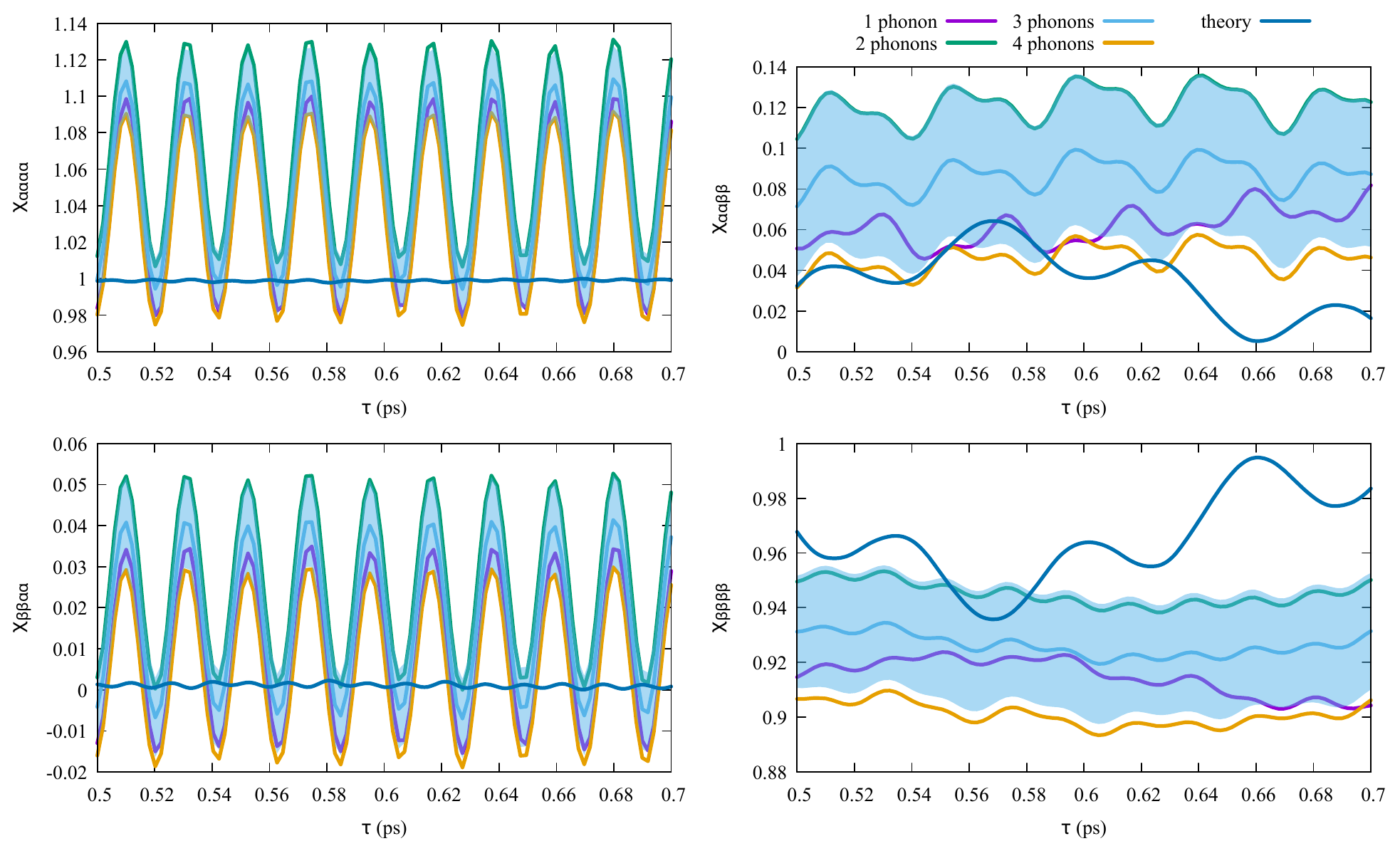}
\caption{Process tensor elements for APC dimers as a function of the total time $\tau = T_1 + T_2 =  t_{\text{P}'}-t_{\text{P}}$, and numbers of phonons. Shown are the results for 1-4 phonons per vibrational model, as well as the theoretically predicted results (dark blue) and the are of one standard deviation around the average for 3 phonons.}
\label{Fig:6}
\end{figure*}

The vibronic progressions for APC overlap significantly as Fig. \ref{Fig:4} illustrates, with the first vibrationally excited state of the $\alpha$-manifold almost degenerate with the vibrational ground state of the $\beta$-manifold. As such, pulse selectivity by tuning the frequency and width of the pulses is difficult to achieve: while possible in principle, the narrowness in frequency would lead (conventionally) to a very long pulse (on the order of several 100 fs) which would mix the natural evolution with processes induced by the continuous action of the pulse, rendering our moot our witnessing of quantum coherence. However, for the single dimer\footnote{Ultrafast single molecule spectroscopy, while challenging, is tractable. Indeed it has provided intriguing insights such as that the coherences in LH2 persist almost ten times longer than the coherences of an isolated fluorophore in a solid state polymer environment~\cite{Brinks2014}. More recently, there have also been room temperature ultrafast nonlinear spectroscopy experiments of single molecules~\cite{Liebel2017}. } we can choose the polarisation of the pulses to be aligned with one transition each, leading to the other transition being dark to that respective pulse (due to the mutual orthogonality of the two transitions). We can thus overcome the restriction imposed by the width of the pulse in frequency. This has consequences though. Starting from Eq. (\ref{Eq:S1}) we can simplify the expression by considering that the pump only excites into a given state, say $\left|i\right\rangle$, while the probe, P$'$, only excites into state $\left|j\right\rangle$. Then,
\begin{equation}
\mathcal{S}(\tau) = \Pi_{f\bar{j}}^{\text{P}'}\Pi_{ig}^{\text{P}}\chi_{\bar{j}\bar{j}ii}(\tau)-\Pi_{gj}^{\text{P}'}\Pi_{ig}^{\text{P}}\chi_{jjii}(\tau) -  \Pi_{gj}^{\text{P}'}\Pi_{ig}^{\text{P}},
\end{equation}
where $\bar{j}$ represents the state that would need to be excited to achieve the transition from $\left|j\right\rangle$ to $\left|f\right\rangle$. For our model dimer we have $i,j\in\left\{\alpha,\beta\right\}$ and we can write four equations which separate into two systems of two equations with two unknowns each. For $j=i$ the pump and probe pulses are identical (P and P), whereas for $j\neq i$ they are the opposites of each other (P and -P). We therefore find,
\begin{equation}
\mathcal{S}^{\text{P},\text{P}}(\tau) = \Pi_{fj}^{\text{P}}\Pi_{ig}^{\text{P}}\chi_{jjii}(\tau) - \Pi_{gi}^{\text{P}}\Pi_{ig}^{\text{P}}\chi_{iiii}(\tau) - \Pi_{gi}^{\text{P}}\Pi_{ig}^{\text{P}},
\label{Eq:SPP}
\end{equation}
\begin{equation}
\mathcal{S}^{-\text{P},\text{P}}(\tau) = \Pi_{fi}^{-\text{P}}\Pi_{ig}^{\text{P}}\chi_{iiii}(\tau) - \Pi_{gj}^{-\text{P}}\Pi_{ig}^{\text{P}}\chi_{jjii}(\tau) - \Pi_{gj}^{-\text{P}}\Pi_{ig}^{\text{P}}.
\label{Eq:SmPP}
\end{equation}
As is shown explicitly in App.~(\ref{App:PolSel}) these two equations will be linearly dependent for perfectly polarisation selective pulses, meaning the corresponding ${\bf M}$ is singular. Indeed, numerical simulations lead to conditions numbers of order 10$^6$ and above. This linear dependence is due to the equivalence of the $\left|g\right\rangle$ to $\left|i\right\rangle$ and $\left|j\right\rangle$ to $\left|f\right\rangle$ transitions, as the transition probability for both transitions will be identical for a given pulse. One might imagine that the symmetry of these transitions can be broken by introducing a biexciton shift, accounting for any interaction between the excitons in $\left|f\right\rangle$. However, as we show in App.~(\ref{App:Biex}), the introduction of this biexciton shift will be identical for the two transitions $\left|i\right\rangle \to \left|f\right\rangle$ and $\left|j\right\rangle\to\left|f\right\rangle$ which does not actually break the symmetry and the equations remain linearly dependent. The central problem therefore remains. By making the pulses perfectly selective we introduce symmetries into the system which lead to linear dependence which cannot be overcome. In fact, even frequency selective pulses lead to the same issues as outlined above. In order to facilitate the inversion a certain amount of cross-talk between the pulses and states seems to be necessary, in direct contradiction to the requirements of our quantum witness.

We next investigate a more realistic scenario by considering an ensemble of dimers randomly oriented with respect to the light pulses. This leads, within the the requirement of near-instantaneous pulses, to excitation of both states with each pulse, i.e., pulse selectivity cannot be perfectly guaranteed generally. This leads to a lifting of the symmetry mentioned above and we have to establish how large the error introduced by the ensuing cross-talk is in our signal expansion and the subsequent inversion.

\subsection{Isotropic Ensemble Spectroscopy}
\label{sec:ens}

We have simulated APC dimers with varying numbers of phonons according to the protocol set out in Sec.~\ref{sec:ExSD}, letting the dimer interact with two pulses of light with defined parameters at defined times. The isotropic averaging was done in a 'brute-force' way: by choosing the angle between both site dipole moments ($40^{\circ}$) the orientation of the dimer was randomised while keeping the polarisation angle between the two pulses (pump and probe) fixed at the magic angle ($\sim 54.7^{\circ}$). %Due to the computational cost involved in these simulations the number of samples varied between the simulations for different number of phonons and qualitative convergence was sought.

Fig. \ref{Fig:6} shows the four process matrix elements as a function of delay time, $\tau$, and for different numbers of phonons. Also shown is the theoretically predicted result from Sec.~(\ref{sec:wit}) and the region of one standard deviation around the curve for 3 phonons (blue shade). 
While we have not presented the 5 phonon results for clarity here, the region of one standard deviation around the 5-phonon curve overlapps with the numerical curves from lower phonon numbers. We can therefore infer that the calculations are essentially converged with 3 phonons. This is in line with the spectrum shown in Fig. \ref{Fig:4}, where we can see an overlap of the one-phonon line with the $\left|\beta\right\rangle$ manifold.

% firstly note that the curve for 5 phonons differs numerically from the area containing the cases for 1 - 4 phonons (which is broadly the area of one standard deviation around the 3 phonon case). We attribute this to the fact that the sample size for 5 phonons was significantly smaller than for the other cases due to computational limitations. 

\begin{figure}[t]
\centering
\includegraphics[width=1\linewidth]{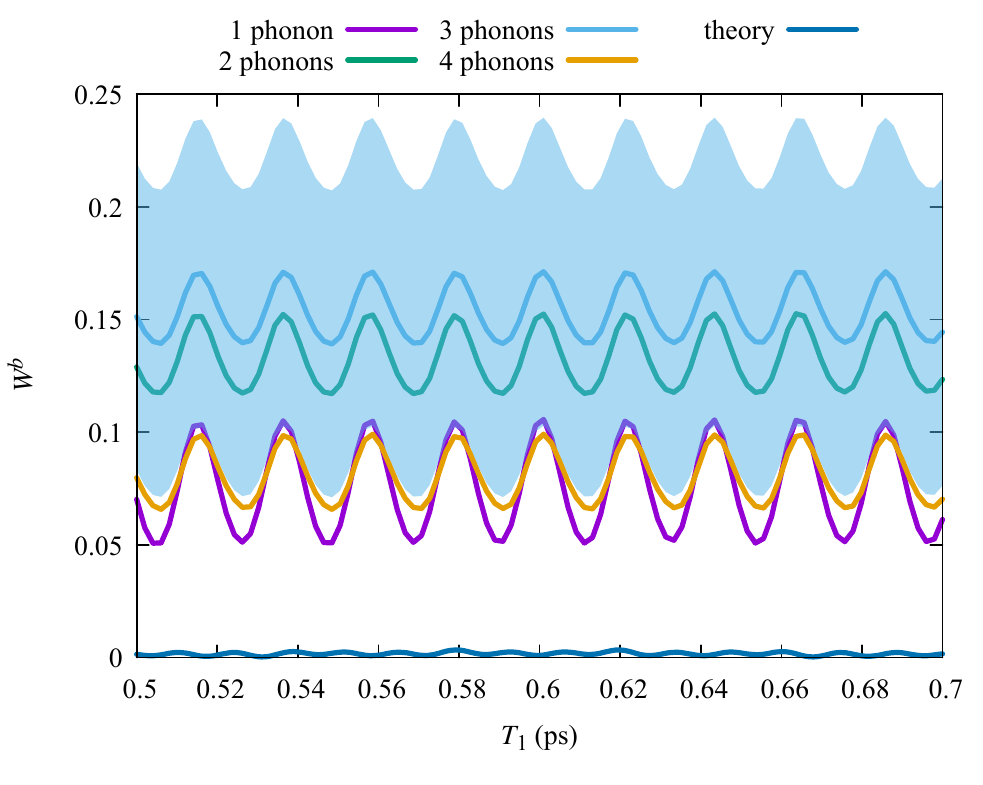}
\caption{NSIT witness $W^b$ for APC, as a function of the interruption time $T_1 = t_{\Gamma} - t_{\text{P}}$, using the results from Fig. \ref{Fig:6} and the theoretical result.}
\label{Fig:7}
\end{figure}

While the curves are broadly within the expected regions ($\chi_{iiii} \simeq 1;\chi_{jjii}\simeq 0$) we can see that they do not agree with the theoretically predicted result. In fact, even qualitatively it is hard to justify any agreement. While it is possible that this is due to non-convergence, we are confident that we can rule out this scenario for the reasons discussed above. We have also calculated the NSIT witness $W^b$ from our numerical simulations and the theoretical prediction in Fig. \ref{Fig:7}. Unsurprisingly, we do not find agreement between simulation and theory here either. 

	As can be seen, the discrepancy is the values of the NSIT witness can be traced back to that in the values of the $\chi$ matrix. Given a molecular system such as APC in our case, the issue arises due to a known trade-off in choosing pulses that are broad enough to span the entire vibrational ladders while also narrow enough to selectively excite the excitons~\cite{Yuen-Zhou:2014ab}. The pulse parameters used in our simulations are laden towards the latter and were chosen after a numerical search providing a well-conditioned ${\bf M}$ matrix. 
	
In the next section, we explore the choice of molecular systems for witnessing quantum coherence using our scheme experimentally.

\subsection{Strong Electronic Coupling}
\label{sec:SEC}

\begin{figure}[t]
\centering
\includegraphics[width=1\linewidth]{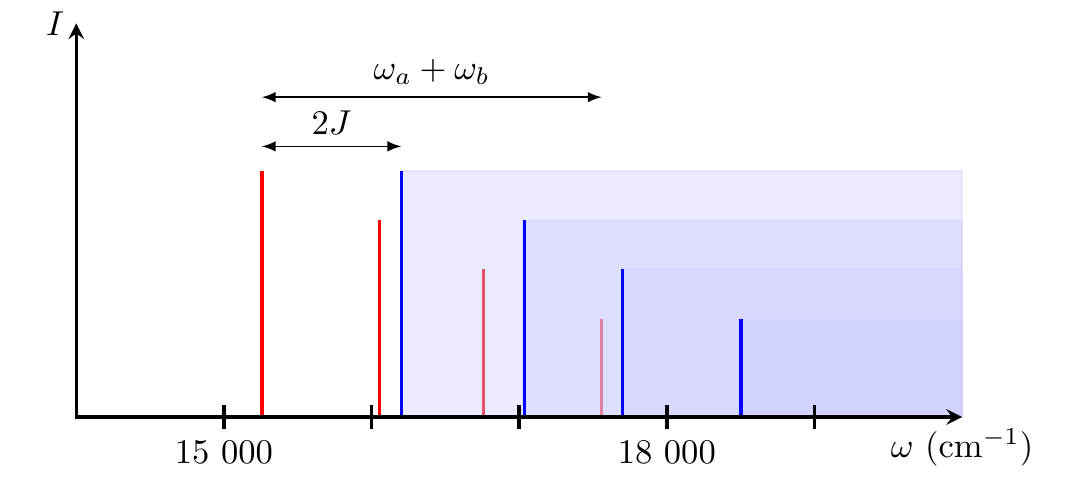}
\caption{Energy scales for APC. Changing $J$ while keeping the vibrational frequencies fixed will shift the two manifold relative to each other, leading to more or less overlap and as a consequence of the targeted pulses to more or less cross-talk between the pulses and manifolds.}
\label{Fig:8}
\end{figure}

To investigate further the disagreement above, we imagine a hypothetical dimer with increasing separation between the vibrational manifolds associated with $\left|\alpha\right\rangle$ and $\left|\beta\right\rangle.$
By truncating the vibrational space of each site to one phonon, we limit the sub-space for each 1EM state to eight states for each excitonic state. In the uncoupled limit ($\hat{H}_{\text{S-B}} = 0$) the four transitions for $\left|\alpha\right\rangle$ are $\varepsilon_{\alpha}^{(0)}, \varepsilon_{\alpha}^{(0)} + \omega_a, \varepsilon_{\alpha}^{(0)} +\omega_b, \varepsilon_{\alpha}^{(0)} + \omega_a + \omega_b$ and similarly for $\left|\beta\right\rangle.$ Each manifold therefore spans an energy range of $\omega_a + \omega_b$, while the energy difference between the vibrational ground states is given by $2J$. Lower cross-talk between the manifolds then amounts to larger values of the quantity 
%we need to have $2J \geq \omega_a + \omega_b$ as in this instance the manifolds are well-separated. By defining,
\begin{equation}
r = \frac{2J}{\omega_a + \omega_b},
\end{equation} 
the ratio between the electronic and vibrational spacing. For $r>1$ the two spectra are non-overlapping in the uncoupled case. Fig. \ref{Fig:8} shows the energy scales involved. For $\hat{H}_{\text{S-B}} \neq 0$ this becomes less-accurate but in the weak-coupling regime is still a useful parameter. We have simulated the hypothetical model for various values of $r$ and compared to the theoretical prediction of our quantum process, where we define the deviation parameter $\sigma$ as
\begin{equation}
\sigma =	\frac{\int_{t_0}^{t_1} \left|\vec{\boldsymbol{\chi}}_{\text{theo}}(t) - \vec{\boldsymbol{\chi}}_{\text{sim}}(t)\right|^2\text{d}t}{\int_{t_{0}}^{t_1} \left|\vec{\boldsymbol{\chi}}_{\text{theo}}(t)\right|^2\text{d}t},
\end{equation}
where $\vec{\chi}_{\text{theo}}$ and $\vec{\chi}_{\text{calc}}$ are the theoretical and calculated reduced process tensors, respectively. $\sigma$ is effectively the average squared deviation over all process tensor elements normalised to the theoretical prediction.

\begin{figure}[h]
\centering
\includegraphics[width=1\linewidth]{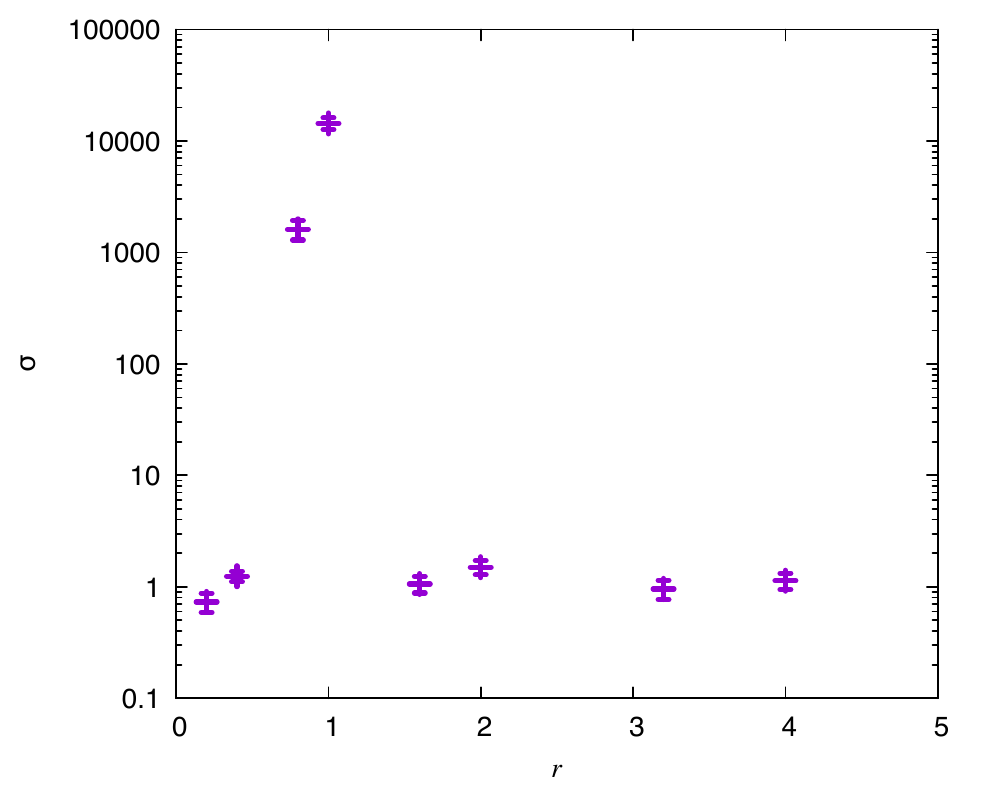}
\caption{Summed squared deviation of the calculated process tensor elements from the theoretical values against the coupling strength parameter $r$. 
	The numerical simulations in this work are performed for $r = 0.14$.}
\label{Fig:9}
\end{figure}

Figure \ref{Fig:9} shows that larger the value of $r$ do not provide radically different values of $\sigma.$ A fuller optimisation could seek to search over both light and matter parameters simultaneously that provide well-conditioned ${\bf M}$  matrices. This would also benefit from extending methods to characterise population transfer~\cite{Dostal2016} and QPT~\cite{Gururangan2019} in EET using 2DES to pump-probe spectroscopy.

%This underlines our results from Sec. 4.2, i.e., the pulses have to be targeting to a reasonable extent, while not being totally selective. Note also that complete separation of the manifolds leads to trivial excited state dynamics.

\section{Conclusions}
\label{sec:conc}

We have presented a quantum witness capable of reporting on the existence of quantum coherences within an electronic system embedded into a vibrational bath, as commonly found in excitonic energy transport in biological systems. The witness relies on \emph{partial} quantum process tomography of the population-to-population elements during the natural evolution of the excited states and comparing a pure classical path-way of transport to the actual process. It may be implemented with four separate two-pulse pump-probe experiments. Any deviation between these two is then attributed to quantum coherent transport and the witness returns a non-zero value. We then proceeded to apply this witness to a coupled dimer system with one vibrational normal mode on each site and simulated the required pump-probe spectra necessary to obtain the required process tensor elements for a single dimer and an isotropic average. Differences from theoretical expectations were discussed in light of the trade-offs involved in the choice of light pulse and matter parameters. Our work shows that unambiguous witnessing of quantum coherence is strongly contingent on performing partial quantum process tomography of a handful of population-to-population elements using two-pulse pump-probe spectroscopy.

Ultrafast pump-probe spectroscopy experiments have recently detected coherent electronic and vibrational oscillations during EET the FMO complex. Using broadband pulses about 12 fs long, they suggested that their detected coherent oscillations comprise at least part of those observed in 2DES experiments~\cite{Maiuri2018}. However, they are unable to confirm the quantum nature of these oscillations or rule our their classical nature. The pump-probe experiments we propose can, using pulses about 100 fs long, albeit for a very different system. Relating them directly to oscillations in 2DES experiments is another question.

That our quantum coherence witness $W^b$ is a faithful reflection of quantum coherence as captured by $R(\rho_{\text{S}})$ rests on certain assumptions. Most significant in the principle of formulating our witness is the Born approximation. We can however still infer the quantum coherence of EET $R(\rho_{\text{S}})$ in case of its invalidity using 
\begin{eqnarray}
R(\rho_{\text{S}}(t_{\Gamma})) & \geq&  2\big(W^b(t_{\Gamma};\tau) - ||\gamma_{\text{SB}}(t_{\Gamma})||_{\text{tr}}  \\ \nonumber
	& & -  ||\tr_{\text{S}}\left[\rho_{\text{SB}}(t_{\Gamma})\right] - \rho_{\text{B}}(t_{\text{P}})||_{\text{tr}}\big),
\end{eqnarray}
if we can lower-bound the violation of the Born approximation and the change in the bath during our spectroscopic protocol. Of course, the system-bath correlations and the bath dynamics have to be small to obtain a non-trivial bound on the excitonic quantum coherence. Recent electronic-vibrational spectroscopic methods~\cite{Oliver:2014aa} may be of use in this regard.

The next significant assumption is the non-invasiveness of the $\Gamma$ operation. In our spectroscopic protocol, this is implemented using light pulses which are invasive. Not only do they not excise the off-diagonal terms instantaneously but also affect the subsequent dynamics of the systems. This can be addressed experimentally~\cite{Leggett:1988aa,Wilde:2012aa,Knee:2016aa} by designing control experiments that capture the invasiveness of $\Gamma$ quantitatively. 

Another assumption in our derivation of the witness $W^b$ is that any coherence within the system is generated only by the natural evolution as the witness itself does not discriminate between possible sources of the coherence. Indeed if $\Gamma$ is implemented exactly as prescribed, this would be our primary challenge in identifying quantum coherence unambiguously. Overcoming this requires temporally narrow pulses to exclude processes induced by the continued interaction with the light pulses and spectrally narrow pulses to avoid creating coherences between excitonic states with the light itself. Transform limited pulses prohibit this. Time-frequency entangled photon-pairs generated by parametric down-conversion (PDC) may allow simultaneous temporal and spectral resolution that is not possible using any `classical' light source, such as a short laser pulse of broadband thermal-like light~\cite{Raymer:2013aa}.

All of the above assumptions are violated to varying degrees in our simulations of EET, as they would be in an actual spectroscopic experiment. These will compromise the unambiguity of our quantum coherence witness, if applied blindly. While none of these assumptions are ever likely to be satisfied exactly, we have discussed how bounding their validity using control and allied characterising experiments can lead towards an unambiguous quantum coherence witness.

\section*{Acknowledgements}
We thank the Engineering and Physical Sciences Research Council, UK (Grant No. EP/K04057X/2) and the Royal Commission for the Exhibition of 1851 for financial support, as well as the CSC, Warwick University, and the Midlands+ HPC for computational support. We also thank Andy Marcus, Jeff Cina, Susana Huelga, Martin Plenio, and Alexandra Olaya-Castro for clarifying discussions and feedback.

%%%END OF MAIN TEXT%%%

%%%REFERENCES%%%
\bibliography{library}

\appendix

\section{Computational Details}
\label{App:Comp}

In order to simulate the spectroscopic experiments we directly integrated the time-dependent Schr\"{o}dinger equation by adding the time-dependent light field described by Eq.~(\ref{Eq:Pulset}) originating from the two pulses in a two-pulse pump-probe experiment and using the resulting Hamiltonian to evolve a state initialised in the true (i.e. electronic and vibrational) ground state. The observable we are interested in is the change of intensity of this light field as it interacts with the sample. As we use a classical light field we have to track these changes indirectly by monitoring the changes in populations of the different states of the electronic system. For instance, a net loss of population in the ground state during one integration time-step corresponds to a loss in intensity of the light pulse due to absorption, while a net-loss in the 2EM shows emission (as we do not consider any electronic states above 2EM). By keeping track of these changes it is possible to simulate a time-resolved signal as is would be recorded in single molecule experiments under the simulated conditions. 

The direct simulation of the two-pulse experiment acting on one molecule requires a careful consideration of the relative phase of the two fields experienced by the molecule. Depending on the spatial positioning the molecule will experience slightly different relative phases from the pulses. While automatically taken care off in real-space simulations, we do not consider the position of the molecule explicitly so the differences in the phase has to be incorporated directly into the simulation. One way would be to introduce a random phase-shift in each simulation and averaging out the relative phase differences in this way. However, computational cost can be reduced by considering the same molecule with two different phase shifts: 0 and $\pi$. In this way we simulate each molecule experiencing an average field instead of averaging over molecules experiencing slightly different phase shifts.

The observable studied in two-pulse pump-probe spectroscopy is the signal due to the excited state. As such the contribution to the signal from the ground state has to be subtracted. To this purpose we simulate the response of the molecule to the probe pulse only. The difference between the phase-averaged signal from the full pump-probe simulation and the probe-only will then give the required time-resolved signal due to the excited state transitions. Integrating these signals lastly gives rise to the integrated signal required for the inversion in Eq. (\ref{Eq:Mat}).

Lastly, to average over the relative orientation of the sample dipole to the light polarisation the latter is kept fixed while the sample is oriented randomly with a fixed angle between the site dipoles in order to simulate a bound dimer system.

\section{Polarisation-Selective Pulses}
\label{App:PolSel}

In the case of single molecule (or dimer) spectroscopy the pulses can be chose such that one transition is dark by ensuring the polarisation is aligned with one and perpendicular to the other transition dipole moment. We thus need to ensure that the system under investigation has orthogonal dipole moments for the two states of the 1EM. We can write,
\begin{equation}
{\boldsymbol \mu}_{x} = \sum_{i} x_{i} {\boldsymbol \mu}_i,
\end{equation}
where $x$ is the energy eigenstate and $x_{i}$ is the wavefunction associated with site $i.$ Then,
\begin{equation}
{\boldsymbol \mu}_{x} \cdot{\boldsymbol \mu}_{x'} = \sum_{ij} x_i x'_j {\boldsymbol \mu}_i \cdot {\boldsymbol \mu}_j.
\end{equation}
For a dimer this can be re-written as,
\begin{equation}
{\boldsymbol \mu}_{\alpha} \cdot {\boldsymbol \mu}_{\beta} = \left[\alpha_a\beta_b  + \alpha_b \beta_a + \alpha_a\beta_a + \alpha_b\beta_b\right]{\boldsymbol \mu}_a\cdot{\boldsymbol \mu}_b.
\end{equation}
For any dimer we find that $\alpha_a = \beta_b$ and $\alpha_b = -\beta_a$ whereby
\begin{equation}
{\boldsymbol \mu}_{\alpha} \cdot {\boldsymbol \mu}_{\beta} = \left[\alpha_a^2 - \alpha_b^2\right]{\boldsymbol \mu}_a\cdot{\boldsymbol \mu}_b.
\end{equation}
In order for the two transition dipole moments to vanish we therefore either need a homodimer ($\alpha_a = \alpha_b, \varepsilon_a = \varepsilon_b$) or we need the two sites constituting the dimer to have orthogonal dipole moments.

This has direct consequences for the calculation of the transition probabilities. Since
\begin{equation}
\Pi_{ij}^p =\left\{\frac{{\boldsymbol \mu}_{ij}\cdot{\bf e}_p}{\hbar}\exp\left[-\frac{\left(\Delta E_{ij} - E_p\right)^2}{2\sigma_p^2}\right]\right\}^2.
\end{equation}
However, if the pulses are in resonance with the transition and the pulses are perfectly selective by polarization then $\Delta E_{ij} - E_p = 0$ and ${\boldsymbol \mu}_{ij} \cdot {\bf e}_p = \mu_{ij}\eta_p$, where $\eta_p$ is the intensity of the pulse $p$. Consequently,
\begin{equation}
\Pi_{ij}^p = \left(\frac{\mu_{ij}\eta_p}{\hbar}\right)^2.
\end{equation}
It then follows, for two perfectly polarization selective pulses that, in order for Eqs. (\ref{Eq:SPP}) and (\ref{Eq:SmPP}) to be linearly independent we require,
\begin{equation}
\mu_{fi}^2\mu_{fj}^2 \neq \mu_{gj}^2\mu_{gi}^2,
\end{equation}
but $\mu_{gi} = \mu_i = \mu_{fj}$ and $\mu_{gj} = \mu_j = \mu_{fi}$ and hence,
\begin{equation}
\mu_i^2\mu_j^2 \neq \mu_i^2\mu_j^2,
\end{equation}
which cannot be true and therefore the inversion fails.

\section{Biexciton Shift}
\label{App:Biex}

The introduction of a biexciton shift will lift the degeneracy of the transitions from the ground state to the 1EM and from the 1EM to the 2EM. for instance, if $E_g = 0$ then $\Delta E_{ig} = E_i$ but $\Delta E_{fj} = E_i + \delta E$, where $\delta E$ quantifies the interaction between two excitations on the dimer. As a direct consequence, the probe pulse won't be resonant with the transition to the 2EM if it is resonant with the transitions to the 1EM and will therefore change the transition probability amplitudes. In order for Eqs. (\ref{Eq:SPP}) and (\ref{Eq:SmPP}) to be linearly independent we require,
\begin{equation}
\Pi_{fi}^{-P}\Pi_{fj}^{P} \neq \Pi_{gj}^{-P}\Pi_{gi}^P.
\end{equation}
Assuming equivalence between transition dipole moments as above and that $\eta_P = \eta_{-P}$ as well as $\sigma_P = \sigma_{-P}$ (i.e. the only adjustable parameters of the pulses are polarization and central frequency) then this reduces to,
\begin{equation}
\left(\Delta E_{fi} - E_{-P}\right)^2 - \left(\Delta E_{gj} - E_{-P}\right)^2 \neq \left(\Delta E_{fj} - E_{P}\right)^2 - \left(\Delta E_{gi} - E_P\right)^2.
\label{Eq:BIS}
\end{equation}
While $E_{gi} = E_P$ and $E_{gj} = E_{-P}$, the biexciton shift will introduce an offset, i.e., $\Delta E_{fj} - E_{P} = \delta E$ and $\Delta E_{fi} - E_{-P} = \delta E$. However, this does not fulfil Eq. (\ref{Eq:BIS}) as it leads to $\delta E \neq \delta E$ and as such the biexciton shift does not lead to linearly dependent equations, causing the inversion to fail, as well.

\end{document}